\documentclass[aps,prd,preprint,groupedaddress]{revtex4}
\usepackage{amsmath}

\newcommand{\bfu}{{\bf u}}
\newcommand{\bfv}{{\bf v}}
\newcommand{\bfw}{{\bf w}}

\newcommand{\bfnabla}{{\mbox{\boldmath $\nabla$}}}

\begin{document}


\title{Quaternionic Madelung Transformation\\ and Non-Abelian Fluid Dynamics}


\author{Peter J. Love, Bruce M. Boghosian}
\affiliation{Department of Mathematics, Tufts University, Medford, MA 02155, U.S.A.}

\date{\today}

\begin{abstract}
In the 1920's, Madelung noticed that if the complex Schroedinger wavefunction is expressed in polar form, $\psi = R e^{i\theta}$, then the quantities $\rho=R^2$ and $\bfu = \bfnabla\theta$ may be interpreted as the hydrodynamic density and velocity, respectively,  of a compressible fluid.  In this paper, we generalize Madelung's transformation to the quaternionic Schroedinger equation. The non-abelian nature of the full $SU(2)$ gauge group of this equation leads to a richer, more intricate set of fluid equations than those arising from complex quantum mechanics. We begin by describing the quaternionic version of Madelung's transformation, and identifying its ``hydrodynamic'' variables. In order to find Hamiltonian equations of motion for these, we first develop the canonical Poisson bracket and Hamiltonian for the quaternionic Schroedinger equation, and then apply Madelung's transformation to derive non-canonical Poisson brackets yielding the desired equations of motion. These are a particularly natural set of equations for a non-abelian fluid, and differ from those obtained by Bistrovic {\it et al.} only by a global gauge transformation~\cite{bib:jackiw2}. Because we have obtained these equations by a transformation of the quaternionic Schroedinger equation, and because many techniques for simulating complex quantum mechanics generalize straightforwardly to the quaternionic case,  our observation leads to simple algorithms for the computer simulation of non-abelian fluids.
\end{abstract}

\pacs{}

\maketitle

\section{Introduction}

Since the $1930$'s it has been known that it is possible to construct quantum mechanics over fields other than the complex numbers. In particular, a detailed and consistent presentation of quaternionic quantum mechanics has been developed and found to have a number of fascinating theoretical properties~\cite{bib:adlerbook}. In the present work, we develop a relationship between quaternionic quantum theory and the simulation of non-relativistic, non-abelian fluid dynamics. Many computer simulation techniques developed for complex quantum mechanics generalize straightforwardly to the quaternionic case, so the observation made in this paper makes possible new simulation strategies for non-abelian fluid dynamics.

The Madelung transformation transforms the complex Schroedinger equation into the continuity and Euler equations describing an inviscid fluid. The magnitude squared of the wavefunction is the mass density, and the gradient of the phase of the wavefunction is the hydrodynamic velocity. The mass continuity equation follows directly from the conservation of probability density, and the momentum equation of motion follows with the identification of a density dependent pressure.

Recently, there has been interest in developing non-abelian fluid dynamical models to describe systems such as quark-gluon plasmas, where one may wish the color degrees of freedom to be retained in the fluid description~\cite{bib:jackiw1,bib:jackiw2,bib:jackiw3}. The color degrees of freedom are associated with the gauge symmetry group of the field; for the example of a quark-gluon plasma this would be the group $SU(3)$ of quantum chromodynamics (QCD). 

A natural question to ask in the context of quaternionic quantum mechanics is: What is the generalization of the Madelung transformation for the quaternionic Schroedinger equation?  As the quaternionic elements are anticommuting, and in particular have the Lie algebra of $SU(2)$, we might plausibly expect the fluid equations arising from such a transformation to be those of a non-abelian fluid.

The remainder of this paper is organized as follows. We first identify the quaternionic generalization of the Madelung transformation, and identify the hydrodynamic variables.  In order to find evolution equations for these quantities we cast the quaternionic Schroedinger equation in Hamiltonian form, defining a canonical Poisson bracket and Hamiltonian for the quaternionic wavefunction. We then use our identification of the quaternionic Madelung  variables to define a non-canonical transformation of our Hamiltonian field theory, obtaining the bracket structure and evolution equations. We compare our approach with that of Jackiw {\it et al.}~\cite{bib:jackiw1,bib:jackiw2, bib:jackiw3}, and briefly present several numerical algorithms by which our equations may be straightforwardly simulated. We close the paper with some conclusions and directions for future work.

\section{Madelung variables}

The identification of the gradient of the phase as the relevant hydrodynamic variable in the complex case is a strong indication that  local gauge symmetry plays an important role in the transformation, and will therefore be useful as a guide in the more difficult quaternionic case.  We first identify the gradient of the phase of the wavefunction, which is identified with the fluid velocity in the complex Madelung transformation, as a pure $U(1)$ gauge. The condition that the fluid velocity be a pure gauge translates to the constraint of vanishing vorticity for the Madelung fluid. We then use this observation to obtain the form of a pure gauge for a quaternionic $SU(2)$ Yang-Mills gauge field. We find three vector fields and show that they obey constraints which are the generalization of vanishing vorticity to the non-abelian case.

To perform the program outlined above we first construct the gauge transformation of a gauge field $A_i$ with arbitrary gauge symmetry group ${\cal G}$. Our gauge field is introduced via the gauge covariant derivative $D_\alpha=\partial_\alpha - A_\alpha$. All physical quantities are invariant under the transformations
\begin{equation}
\psi \rightarrow \psi'=g\psi
\end{equation}
and
\begin{equation}
A_\alpha \rightarrow A_\alpha',
\end{equation}
where $g$ is an element of the gauge symmetry group and the transformation properties of $A_i$ are obtained by requiring
\begin{equation}
\bigl(\partial_\alpha - A'_\alpha\bigr)g\psi = g\bigl(\partial_\alpha- A_\alpha\bigr)\psi,
\end{equation}
which gives in turn
\begin{equation}
A_\alpha'=\bigl(\partial_\alpha g\bigr)g^{-1} + gA_\alpha g^{-1}.
\end{equation}

For complex quantum mechanics coupled to the electromagnetic field the gauge group ${\cal G}$ is $U(1)$, the element $g$ is any unimodular complex number, and the quantity $\bigl(\partial_\mu g\bigr)g^{-1}$ is simply the gradient of the change in phase of the wavefunction. The above transformation must produce no change in electromagnetic field strength $F_{\alpha\beta} = \partial_\alpha A_\beta - \partial_\beta A_\alpha$. It is clear that they do not, as the change in $A_\mu$ takes the form of a similarity transformation and the addition of the pure gauge $\bigl(\partial_\alpha g\bigr)g^{-1}$. The condition that the gauge fields remain unchanged is the elementary condition $\nabla\times\nabla\theta =0$, {\it i.e.}, the constraint that the Madelung fluid has vanishing vorticity.
 
The above discussion motivates a straightforward generalization of the Madelung transformation. The Madelung velocity fields have the form of pure $U(1)$ gauges, and the condition that pure gauges do not contribute to the gauge field strengths translates into the condition of vanishing vorticity for the Madelung velocity field. Turning to the quaternionic case, the gauge group ${\cal G}$ is $SU(2)$, the element $g$ is any unimodular quaternion, and we identify the pure gauges $\bigl(\partial_\alpha g\bigr)g^{-1}$ with the relevant hydrodynamic variables for the quaternionic Madelung transformation.

We write the group element $g$ as
\begin{equation}\label{eq:g}
g=  e^{i\mu}e^{j\theta} e^{k\nu}
\end{equation}
Where $i$, $j$ and $k$ are the quaternionic elements satisfying $i^2=j^2=k^2=ijk=-1$, and $\mu$, $\nu$ and $\theta$ are real angles. Substitution of this form for $g$ into Eq.~(\ref{eq:g}) gives
\begin{equation}
(\nabla g\bigr)g^{-1} = i\nabla \mu + e^{i\mu}j e^{-i\mu} \nabla\theta
+ e^{i\mu}e^{j\theta}k e^{-j\theta}e^{-i\mu}\nabla\nu,\\
\end{equation}
We write
\begin{equation}
(\nabla g\bigr)g^{-1} = i\bfu +j\bfv +k\bfw,
\end{equation}
where
\begin{equation}\label{eq:uvw}
\begin{split}
\bfu  &= \nabla\mu + \cos 2\theta \nabla\nu\\
\bfv  &= \cos 2\mu \nabla\theta
+ \sin 2\theta \sin 2\mu\nabla\nu\\
\bfw  &= \sin 2\mu \nabla\theta
- \sin 2\theta \cos 2\mu\nabla\nu.\\
\end{split}
\end{equation}
The definition of the density is $\rho=\psi\bar\psi$, where $\psi$ is the quaternionic wavefunction $\psi=\sqrt\rho g$ and its quaternionic conjugate $\bar\psi$ is $\psi=\sqrt\rho g^{-1}=\sqrt\rho e^{-k\nu}e^{-j\theta}e^{-i\mu}$.  The quaternionic generalizations of the Madelung variables are then $\rho$, $\bfu$, $\bfv$ and $\bfw$. 

These quaternionic Madelung variables are the pure gauges of an $SU(2)$ Yang-Mills theory. We can see this by showing that the Yang-Mills field strengths are zero. The non-abelian generalization of the Maxwell field strength is
\begin{equation}
F_{\alpha\beta} = \partial_\alpha A_\beta - \partial_\beta A_\alpha  -\bigl[A_\alpha,A_\beta\bigr]
\end{equation}
where $A_\alpha$ are quaternionic gauge fields. Sufficient and necessary conditions for these field strengths to be zero are
\begin{equation}
\label{eq:curls}
\begin{split}
\nabla \times \bfu &= 2 \bfv\times\bfw\\
\nabla \times \bfv &= 2 \bfw\times\bfu\\
\nabla \times \bfw &= 2\bfu\times\bfv\\
\end{split}
\end{equation}
It may be readily verified that these equations are satisfied by $\bfu$, $\bfv$ and $\bfw$, and hence are the quaternionic generalization of vanishing vorticity in the complex Madelung transformation. We now obtain the equations of motion for $\bfu$, $\bfv$ and $\bfw$.

\section{Equations of Motion}

In order to obtain the equations of motion for $\bfu, \bfv$ and $\bfw$ we first obtain a canonical Hamiltonian form for the quaternionic Schroedinger equation. To do this we need to define a Hamiltonian and a Poisson bracket. Defining the Hamiltonian is straightforward, as it follows directly from the definition for the complex case,
\begin{equation}
\begin{split}
H &= \frac{1}{2}\int \bar\psi\hat H\psi d^3 z=\frac{1}{2}\int\biggl(-\frac{1}{2}\bar\psi\nabla^2\psi + V\bar\psi\psi\biggr)  d^3z\\
&= \frac{1}{2}\int \biggl( \frac{|\nabla\rho|^2}{8\rho}+\frac{\rho}{2}(\bfu^2 +\bfv^2 +\bfw^2)+ V\rho\biggr)\; d^3x,
\end{split}
\end{equation}
where we use units such that $\hbar = m = 1$ throughout. By the argument presented in Appendix~\ref{sec:app1} we define the bracket 
\begin{equation}
\label{eq:bracket}
\begin{split}
\{Q({\bf x}),P({\bf y})\} &=\int\biggl(\frac{\delta Q({\bf x})}{\delta a({\bf z})}\frac{\delta P({\bf y})}{\delta b({\bf z})}-\frac{\delta Q({\bf x})}{\delta b({\bf z})}\frac{\delta P({\bf y})}{\delta a({\bf z})}\biggr)d^3z\\
&+\int\biggl(\frac{\delta Q({\bf x})}{\delta d({\bf z})}\frac{\delta P({\bf y})}{\delta c({\bf z})}-\frac{\delta Q({\bf x})}{\delta c({\bf z})}\frac{\delta P({\bf y})}{\delta d({\bf z})}\biggr)d^3z,
\end{split}
\end{equation}
where $a$, $b$ , $c$ and $d$ are the quaternionic components of the wavefunction $\psi =a + ib + jc + kd$.  The above Poisson bracket and Hamiltonian reproduce the quaternionic Schroedinger equation. The definitions Eq.~(\ref{eq:uvw}) are then a non-canonical transformation of this Hamiltonian field theory. The bracket structure of the quaternionic Madelung equations follows by lengthy but straightforward algebra. We spare the reader the details and present the noncanonical brackets amongst $\rho$, $\bfu$, $\bfv$ and $\bfw$ in Appendix~\ref{sec:app2}.  Using the above Hamiltonian, the equation of motion for $\rho$ is then
\begin{equation}
\frac{\partial \rho}{\partial t} + \nabla\cdot\bigl(\rho \bfu\bigr) = 0.
\end{equation}
This is the continuity equation, and if $\rho$ is identified as the singlet density, this equation identifies $\bfu$ as the singlet velocity. The equation of motion for $\bfu$ is
\begin{equation}
\label{eq:ueq}
\begin{split}
\frac{\partial \bfu}{\partial t}&=-\nabla V'-\frac{1}{2}\nabla(\bfu^2 +\bfv^2 +\bfw^2)
-\frac{\bfv}{\rho}\nabla\cdot(\rho\bfv)-\frac{\bfw}{\rho}\nabla\cdot(\rho\bfw),
\end{split}
\end{equation}
where $V'$ is the density-dependent potential (also called the Bohmian potential) which also appears in the complex Madelung transformation,
\begin{equation}
V' = V-\frac{1}{4\sqrt{\rho}}\nabla\cdot\biggl(\frac{\nabla\rho}{\sqrt{\rho}}\biggr).
\end{equation} 
Note that the curl of Eq.~(\ref{eq:ueq}) is generally nonvanishing.  Since $\bfu$ is identified as the singlet density, this means that the quaternionic Madelung fluid has non-zero vorticity. This can be clearly seen by inspecting the $\{\bfu({\bf x}),\bfu({\bf y})\}$ bracket, which has the correct form for rotational flow. The equations for $\bfv$ and $\bfw$ are
\begin{equation}
\label{eq:veq}
\begin{split}
\frac{\partial\bfv}{\partial t}&=\frac{1}{2}\nabla\biggl[\frac{\nabla\cdot(\rho\bfw)}{\rho}\biggr] - 2\bfw\biggl[V'({\bf x})+\frac{1}{2}(\bfu^2 +\bfv^2 +\bfw^2)\biggr]+\frac\bfu{\rho}\nabla\cdot(\rho\bfw)
\end{split}
\end{equation}
and
\begin{equation}
\label{eq:weq}
\begin{split}
\frac{\partial\bfw}{\partial t}&=-\frac{1}{2}\nabla\biggl[\frac{\nabla\cdot(\rho\bfv)}{\rho}\biggr] + 2\bfv\biggl[V'({\bf x})+\frac{1}{2}(\bfu^2 +\bfv^2 +\bfw^2)\biggr]+\frac\bfu{\rho}\nabla\cdot(\rho\bfv).
\end{split}
\end{equation}
One immediately notices that the equations for $\bfv$ and $\bfw$ are invariant under the duality transformation
\begin{equation}
\label{eq:duality}
\begin{pmatrix}
\bfv'\\
\bfw'\\
\end{pmatrix}
= \begin{pmatrix}
\cos\theta & -\sin\theta\\
\sin\theta & \cos\theta\\
\end{pmatrix}
\begin{pmatrix}
\bfv\\
\bfw\\
\end{pmatrix}
\end{equation}
This duality symmetry is a consequence of the global gauge invariance of the quaternionic Schroedinger equation. This global invariance is discussed in detail in Appendix~\ref{sec:app3}.

\section{Non-abelian fluid dynamics}

Recently there has been interest in constructing non-abelian fluid dynamics and fluid magnetohydrodynamics in order to simulate the long-wavelength degrees of freedom in a quark-gluon plasma. In particular, Jackiw {\it et al.} have described several models for the relativistic dynamics of such a system~\cite{bib:jackiw1,bib:jackiw3}. Those authors have also presented a non-relativistic multicomponent wavefunction very similar to the quaternionic wavefunction considered here~\cite{bib:jackiw2}. In fact, as we shall show, their model differs from the quaternionic Madelung transformation we have investigated only by a global gauge of the type described in detail in Appendix~\ref{sec:app3}. In this section we compare their multicomponent wave equation with our approach and give the global gauge transformation relating their work to our own. 

Bistrovic {\it et al.} consider a two component complex wave equation,
\begin{equation}\label{jackse}
i\: \frac{d\psi}{dt} = -\frac{1}{2} \nabla^2 \psi.
\end{equation}
Those authors consider the group $SU(2)$ and the fundamental representation: $T^a={\sigma^a}/{(2i)}$, $\{T^a,T^b\}=-\delta^{ab}/2$, where $\sigma^a, a=1,2,3$ are the Pauli matrices. The $T^a$'s here are simply a two dimensional complex representation of the quaternionic elements
\begin{equation}
\mbox{$T^1 = \frac{i}{2}$}, \mbox{$T^2 = \frac{j}{2}$}, \mbox{$T^3 = \frac{k}{2}$}.
\end{equation}
Those authors define the conserved singlet and color current;
\begin{equation}
j^\mu=(c \rho, {\bf j})\;,\qquad \rho
	= \psi^\dagger \psi\;, \qquad
{\bf j} = \mathrm{Im}\, 
	\psi^\dagger \nabla \psi\;.
\end{equation}
\begin{equation}
J^\mu_a=(c \rho_a, {\bf J}_a)\;,\qquad \rho_a
	= i \psi^\dagger T^a \psi\;, \qquad
{\bf J}_a = \mathrm{Re}\,\psi^\dagger T^a 
	\nabla \psi\;.
\end{equation}
and the wavefunction,
\begin{equation}
\psi=\sqrt{\rho} g {\bf A}
\end{equation}
where $\rho$ is the scalar $\psi^\dagger\psi$, where $\dagger$ denotes hermitian conjugation, $g$ is a
group element, and ${\bf A}$ is an arbitrary constant vector taken to be
\begin{equation}
{\bf A}=
\begin{pmatrix}
1\\0\\
\end{pmatrix}
\end{equation} 
in which case $i{\bf A}^\dagger T^a {\bf A}= \delta^{a3}/2$.  The singlet density is $\rho$,
while the singlet current ${\bf j}$ is
\begin{equation}
{\bf j}=\rho{\bf p}\;, \qquad {\bf p}\equiv 
	-i {\bf A}^\dagger  \:g^{-1}\nabla g\:{\bf A}
\;.
\end{equation}
The key relation connecting our approach to Jackiw's is:
\begin{eqnarray}
g^{-1}\nabla g &\equiv& -2{\bf p}^a T^a\\
{\bf p}&=&{\bf p}^a t^a,
\end{eqnarray} 
where $t^a/2=iu^\dagger T^a u= \delta^{a3}/2$, and hence
\begin{equation}
{\bf p}={\bf p}^a t^a= {\bf p}^3.
\end{equation} 
It follows that the singlet velocity in~\cite{bib:jackiw2} is equated to the negative of third quaternionic component of the wavefunction in our approach. His multicomponent wavefunction is therefore equivalent to a quaternionic Schroedinger equation
\begin{equation}
-k\frac{d\psi}{dt} = -\frac{1}{2} \nabla^2 \psi.
\end{equation}
This quaternionic Schroedinger equation is related to the one considered in this paper by a one-parameter family of global gauge transformations, described in detail in Appendix~\ref{sec:app3}.

The color and singlet currents identified in~\cite{bib:jackiw2} may be obtained from the fields $\bfu$, $\bfv$, and $\bfw$ defined for our quaternionic Madelung equation. The quaternionic Madelung transformation we have described is a particularly simple implementation of this type of non-abelian fluid model, but its usefulness is limited as the approach cannot be extended to groups other than $SU(2)$. The quaternionic approach has some advantages in terms of  ease of numerical simulation, which we describe below.

\section{Simulation}

The Schroedinger equation may be simulated by a variety of methods. We briefly describe three methods, all of which may be trivially generalized to the simulation of quaternionic quantum mechanics.  The first method is operator splitting, or its higher order variant, the symplectic integrator method.  This method is perfectly unitary, can be generalized to three dimensions, and can be generalized to high orders of accuracy~\cite{bib:opsplit1,bib:opsplit2,bib:opsplit3,bib:opsplit4,bib:opsplit5,bib:opsplit6}. The second-order spatial derivatives required in the operator splitting method may be avoided by using a quantum lattice-Boltzmann method (QLBE)~\cite{bib:qlbe1,bib:qlbe2, bib:qlbe3}. The QLBE is unconditionally stable, perfectly unitary, and offers considerable advantages for parallel computing applications. 

Closely related to the quantum lattice Boltzmann approach is the quantum lattice-gas model. These algorithms were originally developed for simulation of the Dirac equation in one spatial dimension~\cite{bib:meyer1,bib:meyer2,bib:meyer3,bib:meyer4}. The non-relativistic quantum lattice gas was subsequently developed for the simulation of many-body quantum mechanics in arbitrary numbers of dimensions~\cite{bib:bogwash}. The quantum lattice-gas algorithm for a single nonrelativistic particle provides a tractable technique for the simulation of the quaternionic Schroedinger equation.

It should be noted that the quantum lattice gas algorithm (QLGA) of~\cite{bib:bogwash} was developed in order to find a quantum algorithm for Schroedinger's equation. The QLGA provides a method of simulating quantum mechanics on a quantum computer with an exponential enhancement in the performance. Minor modifications to this QLGA should also provide a method of simulating the quaternionic Madelung equations derived above on a (complex) quantum computer with a similar exponential speedup.

\section{Conclusions}

We have obtained the quaternionic generalization of the Madelung transformation. We have identified the generalizations of the hydrodynamic  fields and obtained their evolution equations by a non-canonical transformation of the Hamiltonian field theory for the quaternionic Schroedinger equation. 

The equations we have obtained also possess a limit which may be of interest. The choice of arbitrarily large potential may always make the second terms in Eqs.~(\ref{eq:veq}) and (\ref{eq:weq}) dominant. In this case a solution which is constant in space and oscillatory in time will be a good first-order approximation to the behavior of $\bfv$ and $\bfw$. This suggests that there may exist an approximate treatment of these equations in which $\bfv$ and $\bfw$ are treated as ``fast'' variables which may be averaged away. This would be analogous to the gyro-averaged description of MHD, in which the fast gyrofrequency motions are averaged away and only the slow motion of the gyro centers is retained in the reduced dynamics~\cite{bib:gyroav1,bib:gyroav2,bib:gyroav3}. The Hamiltonian approach taken in this paper provides a good starting point for the derivation of such an averaged treatment.

The Madelung transformation of complex quantum mechanics seems intially computationally promising, as it shows us how to simulate the nonlinear Euler equations by simulating the linear Schroedinger equation. However, irrotational inviscid fluid flow is not a problem of sufficient interest to merit serious computational investigation. The quaternionic Madelung transformation gives rise to equations of considerably more interest, as they describe the (rotational) flow of a non-abelian fluid.  However, the constraints in Eq.~(\ref{eq:curls}) are the restriction on the non-abelian fluid, analogous to vanishing vorticity in the abelian case. This suggests that there exists a generalization of the equations obtained here to non-abelian flow fields which do not satisfy the constraints of Eq.~(\ref{eq:curls}). It is possible that these general non-abelian flow equations will contain as much additional physics as does rotational Euler flow compared with irrotational Euler flow. 

A natural question arises as to whether one may generalize this approach to groups other than $SU(2)$ and fields other than the quaternions. We may answer this immediately in the negative: There are no division algebras other than the reals, the complex numbers, the quaternions and the the octonions~\cite{bib:quatbook}. The octonions form a division algebra, but are not associative, and this precludes the formulation of an octonionic quantum mechanics~\cite{bib:adlerbook}.

Finally, we have noted that the computational tractability of the complex Schroedinger  equation is shared by its quaternionic generalization, and so we anticipate that extensive simulation of the quaternionic Schroedinger equation by any one of the methods described above will yield results of interest to the particle physics community.

\section*{Acknowledgements}

BMB was supported in part by the U.S. Air Force Office of Scientific Research under grant number F49620-01-1-0385. PJL was supported by the DARPA QuIST program under AFOSR grant number F49620-01-1-0566.

\appendix

\section{Quaternionic Poisson Bracket
\label{sec:app1}}

If we write the Schroedinger equation as a pair of real equations for the real and imaginary parts of the wavefunction $\psi = a+ib$ we obtain:
\begin{equation}
\frac{d}{dt} \begin{pmatrix}
a\\
b\\
\end{pmatrix}
= \begin{pmatrix}
0 & \hat H\\
- \hat H & 0\\
\end{pmatrix}
\begin{pmatrix}
a\\
b\\
\end{pmatrix}
\end{equation}
where $\hat H = -\frac{1}{2}\nabla^2 + V$. Note that the above construction does not enable us to construct quantum mechanics over the reals. The symplectic matrix in the above equation does not have real eigenvalues and so there are no energy eigenstates in real quantum mechanics~\cite{bib:adlerbook}. However, the form of this equation makes it immediately obvious that the Schroedinger equation may be described by a canonical Hamiltonian field theory in which the real and imaginary parts of the equation are canonically conjugate.  The definition of the bracket follows automatically,
\begin{equation}
\begin{split}
\{Q({\bf x}),P({\bf y})\} &=\int\biggl(\frac{\delta Q({\bf x})}{\delta a({\bf z})}\frac{\delta P({\bf x})}{\delta b({\bf z})}-\frac{\delta Q({\bf x})}{\delta b({\bf z})}\frac{\delta P({\bf x})}{\delta a({\bf z})}\biggr)d^3z\\
\end{split}
\end{equation}
The quaternionics have a four dimensional real representation, which enables us to write the quaternionic Schroedinger equation as four real equations for the four components of the wavefunction $\psi=a+ib+jc+kd$,
\begin{equation}
\frac{d}{dt} \begin{pmatrix}
a\\
b\\
c\\
d 
\end{pmatrix}
=  \begin{pmatrix}
0 & \hat H & 0 & 0 \\
-\hat H & 0 & 0 & 0\\
0 & 0 & 0 & -\hat H\\
0 & 0 & \hat H & 0
\end{pmatrix}
\begin{pmatrix}
a\\
b\\
c\\
d 
\end{pmatrix}
\end{equation}
Again, we see that these equations have a symplectic structure where $a$ and $b$ are canonically conjugate, and so are $d$ and $c$. The definition of the bracket, Eq.~(\ref{eq:bracket}), again follows automatically.

\section{Poisson Bracket structure\label{sec:app2}}

Our quaternionic Madelung variables have the following bracket structure:
\begin{equation}
\begin{split}
\{\rho({\bf x}), w_i({\bf y})\}&=-4\delta({\bf x}-{\bf y})v_i({\bf x})\\
\{\rho({\bf x}), v_i({\bf y})\}&=4\delta({\bf x}-{\bf y})w_i({\bf x})\\
\{\rho({\bf x}), u_i({\bf y})\}&=-2\delta_i'({\bf x}-{\bf y})\\
\end{split}
\end{equation}
\begin{equation}
\begin{split}
\{u_i({\bf x}), u_j({\bf y})\}&=\frac{2\delta({\bf x}-{\bf y})}{\rho({\bf x})}\bigl(\partial_i u_j({\bf x})-\partial_j u_i({\bf x})\bigr)\\
\{u_i({\bf x}), v_j({\bf y})\}&=-\frac{2\delta_j'({\bf x}-{\bf y})v_i({\bf x})}{\rho({\bf x})}
+\frac{4\delta({\bf x}-{\bf y})}{\rho({\bf x})}\biggl[\frac{1}{2}\bigl(\partial_iv_j({\bf x})-\partial_jv_i({\bf x})\bigr) -u_i({\bf x})w_j({\bf x})\biggr]\\
\{ u _i({\bf x}),w_j({\bf y})\}&=-\frac{2\delta_j'({\bf x}-{\bf y})w_i({\bf x})}{\rho({\bf x})}-\frac{4\delta({\bf x}-{\bf y})}{\rho({\bf x})}\biggl[\frac{1}{2}(\partial_jw_i({\bf x})-\partial_iw_j({\bf x})) -v_j({\bf x})u_i({\bf x})\biggr]\\
\{v_i({\bf x}), u_j({\bf y})\}&=\frac{2\delta_i'({\bf y}-{\bf x})v_j({\bf y})}{\rho({\bf y})}
-\frac{4\delta({\bf y}-{\bf x})}{\rho({\bf y})}\biggl[\frac{1}{2}\bigl(\partial_jv_i({\bf y})-\partial_iv_j({\bf y})\bigr) -u_j({\bf y})w_i({\bf y})\biggr]\\
\{v_i({\bf x}), v_j({\bf y})\}&=-\frac{2\delta_i'({\bf y}-{\bf x})u_j({\bf y})}{\rho({\bf y})}+\frac{2\delta_j'({\bf x}-{\bf y})u_i({\bf x})}{\rho({\bf x})}\\
&-\frac{4\delta({\bf y}-{\bf x})}{\rho({\bf x})}\biggl[\frac{1}{2}\biggl(\partial_iu_j({\bf x})-\partial_ju_i({\bf x})\biggr)+v_i({\bf x})w_j({\bf x})-v_j({\bf x})w_i({\bf x})\biggr]\\
\{v_i({\bf x}), w _j({\bf y})\}&=\frac{\delta_{ij}''({\bf x}-{\bf y})}{\rho({\bf x})}-\frac{\delta_j'({\bf x}-{\bf y})\partial_i\rho({\bf x})}{\rho({\bf x})^2}
-4\delta({\bf y}-{\bf x})\frac{u_j({\bf x})u_i({\bf x})}{\rho({\bf x})}\\
\{w_i({\bf x}), u _j({\bf y})\}&=\frac{2\delta_i'({\bf y}-{\bf x})w_j({\bf y})}{\rho({\bf y})}+\frac{4\delta({\bf x}-{\bf y})}{\rho({\bf x})}\biggl[\frac{1}{2}(\partial_iw_j({\bf x})-\partial_jw_i({\bf x})) -v_i({\bf x})u_j({\bf x})\biggr]\\
\{w_i({\bf x}), v _j({\bf y})\}&=-\frac{\delta_{ij}''({\bf y}-{\bf x})}{\rho({\bf y})}+\frac{\delta_i'({\bf y}-{\bf x})\partial_j\rho({\bf y})}{\rho({\bf y})^2}
+4\delta({\bf x}-{\bf y})\frac{u_i({\bf x})u_j({\bf x})}{\rho({\bf x})}\\
\{w_i({\bf x}), w_j({\bf y})\}&=-2\frac{\delta_i'({\bf y}-{\bf x})}{\rho({\bf y})}u_j({\bf y})+2\frac{\delta_j'({\bf x}-{\bf y})}{\rho({\bf x})}u_i({\bf x})\\
&+4\delta({\bf y}-{\bf x})\frac{1}{\rho({\bf y})}\biggl[\frac{1}{2}(\partial_iu_j({\bf y})-\partial_ju_i({\bf y}))+v_i({\bf y})w_j({\bf y})-v_j({\bf y})w_i({\bf y})\biggr]\\
\end{split}
\end{equation}
\section{Global gauge symmetry\label{sec:app3}}
We obtained our quaternionic Madelung variables by considering the local gauge invariance properties of our Schroedinger equation coupled to a quaternionic gauge field. The observables of quaternionic quantum mechanics posess global $SU(2)$ gauge invariance, and this invariance will lead to a set of symmetry properties for the quaternionic Madelung equations. The Schroedinger equation is
\begin{equation}\label{eq:gg1}
i\frac{\partial \psi}{\partial t} =\hat H\psi
\end{equation}
Under the global gauge transformation $\psi \rightarrow \psi' = \phi\psi$ this becomes:
\begin{equation}\label{eq:gg2}
\phi^{-1}i\phi\frac{\partial \psi}{\partial t} =\hat H \psi
\end{equation}
If $\phi$ is a unimodular quaternion then the transformation $\phi^{-1}Q\phi$, where $Q$ is any quaternion, is an automorphism of the quaternions, and $\phi^{-1}Q\phi$ is also a unimodular quaternion. The equivalent global gauge transformation applied to the complex Schroedinger equation also induces an automorphism of the complex numbers, but in that case it is the trivial automorphism $\forall \theta : e^{-i\theta}Ze^{i\theta} =Z $. The choice of $i$ in Eq.~(\ref{eq:gg1}), which we inherited from the complex Schroedinger  equation, constitutes a global gauge choice. We could equally well write the quaternionic Schroedinger equation replacing $i$ by $j$, or $k$, or indeed any pure imaginary unimodular quaternion. We now consider how such global gauge choices affect our quaternionic Madelung equations.

Our Madelung variables are defined by:
\begin{equation}
(\nabla g\bigr)g^{-1} = i\bfu +j\bfv +k\bfw,
\end{equation}
applying a global gauge transformation to this quantity gives:
\begin{equation}
\phi^{-1}(\nabla g\bigr)g^{-1} \phi= \phi^{-1}i\phi\bfu +\phi^{-1}j\phi\bfv +\phi^{-1}k\phi\bfw,
\end{equation}
Taking the quaternion automorphism defined by:
\begin{equation}
f_p = \bar \phi e_p\phi.
\end{equation}
Where $e_p = i,j,k, p=1,2,3$ are our original quaternionic elements and $f_q$ are our new quaternionic elements. We express the new quaternionic elements in terms of the old:
\begin{equation}
f_q=a_{qp} e_p
\end{equation}
We can determine the properties of the coefficients $a_{qp}$ from the algebra obeyed by both sets of quaternionic elements:
\begin{equation}
\begin{split}
e_qe_r &= -\delta_{qr} + \epsilon_{qrs}e_s\\
\end{split}
\end{equation}
Substituting:
\begin{equation}
\begin{split}
a_{mp}a_{nq}\biggl(-\delta_{pq} +\epsilon_{pqs}e_s\biggr) &= -\delta_{mn} +\epsilon_{mno}a_{or}e_r\\
\end{split}
\end{equation}
Where $\epsilon_{mno}$ is the completely antisymmetric Levi-Civita tensor. Separating the coefficients of the quaternionic elements $e_s$ gives:
\begin{equation}
\begin{split}
a_{rp}a_{sq} \delta_{pq}&= \delta_{rs}\\
\bigl(a_{rp}a_{sq} \epsilon_{pqr} &-\epsilon_{rsl}a_{lr}\bigr)e_m=0\
\end{split}
\end{equation}
These equations show that the matrix $a_{pq}$ is orthogonal with unit determinant. The global gauge transformation induces a linear transformation (actually a proper rotation) of our Madelung variables,
\begin{equation}
\begin{split}
\bfu'& = a_{11}\bfu + a_{21}\bfv + a_{31}\bfw\\
\bfv'& = a_{12}\bfu + a_{22}\bfv + a_{32}\bfw\\
\bfw'& = a_{13}\bfu + a_{23}\bfv + a_{33}\bfw.
\end{split}
\end{equation} 
The Madelung transformation therefore applies to a three-parameter family of quaternionic Schroedinger equations, and the results of simulating any member of this family may be related to the results of simulating any other member by a global gauge transformation. As a specific example of this we construct the transformation relating the quaternionic Schroedinger equations:
\begin{equation}\label{eq:qse}
i \frac{d\psi}{dt} = -\frac{1}{2} \nabla^2 \psi
\end{equation}
and
\begin{equation}
-k \frac{d\psi}{dt} = -\frac{1}{2} \nabla^2 \psi. 
\end{equation}
The two equations are related by the global gauge transformation $\psi' \leftarrow \phi\psi$ where $\phi = e^{i\theta}\bigl(1-j\bigr)/\sqrt{2}$, and the matrix $a_{pq}$ is
\begin{equation}
a_{pq} =
\begin{pmatrix}
0&0&-1\\
0&-1&0\\
-1&0&0\\
\end{pmatrix}
\end{equation}
This is the global gauge transformation relating the quaternionic Schroedinger equation considered here to the multicomponent non-abelian Schroedinger equation considered by Bistrovic {\it et al.}~\cite{bib:jackiw2}. 

The gauge transformations relating the two quaternionic Schroedinger equations above form a one parameter family. There is always a one-parameter subgroup of the full gauge group which leaves any quaternionic Schroediger equation identically invariant. For the Schroedinger equation, Eq.~(\ref{eq:qse}), the elements of this group may be written $g= e^{i\theta/2}$. This subgroup induces the transformations
\begin{equation}
a_{pq} =
\begin{pmatrix}
1&0&0\\
0&\cos{\theta}&-\sin{\theta}\\
0&\sin{\theta}&\cos{\theta}\\
\end{pmatrix}
\end{equation}
This is the set of global gauge transformations which give rise to the duality symmetry of $\bfv$ and $\bfw$ noted in Eq.~(\ref{eq:duality}).



\begin{thebibliography}{22}
\expandafter\ifx\csname natexlab\endcsname\relax\def\natexlab#1{#1}\fi
\expandafter\ifx\csname bibnamefont\endcsname\relax
  \def\bibnamefont#1{#1}\fi
\expandafter\ifx\csname bibfnamefont\endcsname\relax
  \def\bibfnamefont#1{#1}\fi
\expandafter\ifx\csname citenamefont\endcsname\relax
  \def\citenamefont#1{#1}\fi
\expandafter\ifx\csname url\endcsname\relax
  \def\url#1{\texttt{#1}}\fi
\expandafter\ifx\csname urlprefix\endcsname\relax\def\urlprefix{URL }\fi
\providecommand{\bibinfo}[2]{#2}
\providecommand{\eprint}[2][]{\url{#2}}

\bibitem[{\citenamefont{Bistrovic et~al.}(2002)\citenamefont{Bistrovic, Jackiw,
  Li, Nair, and Pi}}]{bib:jackiw2}
\bibinfo{author}{\bibfnamefont{B.}~\bibnamefont{Bistrovic}},
  \bibinfo{author}{\bibfnamefont{R.}~\bibnamefont{Jackiw}},
  \bibinfo{author}{\bibfnamefont{H.}~\bibnamefont{Li}},
  \bibinfo{author}{\bibfnamefont{V.~P.} \bibnamefont{Nair}}, \bibnamefont{and}
  \bibinfo{author}{\bibfnamefont{S.-Y.} \bibnamefont{Pi}},
  \bibinfo{journal}{hep-th/0210143}  (\bibinfo{year}{2002}).

\bibitem[{\citenamefont{Adler}(1995)}]{bib:adlerbook}
\bibinfo{author}{\bibfnamefont{S.~L.} \bibnamefont{Adler}},
  \emph{\bibinfo{title}{Quaternionic quantum mechanics and quantum field
  theory}} (\bibinfo{publisher}{Oxford University Press},
  \bibinfo{year}{1995}).

\bibitem[{\citenamefont{Jackiw et~al.}(2000)\citenamefont{Jackiw, Nair, and
  Pi}}]{bib:jackiw1}
\bibinfo{author}{\bibfnamefont{R.}~\bibnamefont{Jackiw}},
  \bibinfo{author}{\bibfnamefont{V.~P.} \bibnamefont{Nair}}, \bibnamefont{and}
  \bibinfo{author}{\bibfnamefont{S.-Y.} \bibnamefont{Pi}},
  \bibinfo{journal}{Phys. Rev. D} \textbf{\bibinfo{volume}{62}},
  \bibinfo{pages}{085018} (\bibinfo{year}{2000}).

\bibitem[{\citenamefont{Jackiw}(2002)}]{bib:jackiw3}
\bibinfo{author}{\bibfnamefont{R.}~\bibnamefont{Jackiw}},
  \emph{\bibinfo{title}{Lectures on Fluid Dynamics}}
  (\bibinfo{publisher}{Springer}, \bibinfo{year}{2002}).

\bibitem[{\citenamefont{Chin and Chen}(2001)}]{bib:opsplit1}
\bibinfo{author}{\bibfnamefont{S.~A.} \bibnamefont{Chin}} \bibnamefont{and}
  \bibinfo{author}{\bibfnamefont{C.~R.} \bibnamefont{Chen}},
  \bibinfo{journal}{J. Chem. Phys.} \textbf{\bibinfo{volume}{114}},
  \bibinfo{pages}{7338} (\bibinfo{year}{2001}).

\bibitem[{\citenamefont{Feit et~al.}(1982{\natexlab{a}})\citenamefont{Feit,
  Fleck, and Steiger}}]{bib:opsplit2}
\bibinfo{author}{\bibfnamefont{D.}~\bibnamefont{Feit}},
  \bibinfo{author}{\bibfnamefont{J.~A.} \bibnamefont{Fleck}}, \bibnamefont{and}
  \bibinfo{author}{\bibfnamefont{A.}~\bibnamefont{Steiger}},
  \bibinfo{journal}{J. Comput. Phys.} \textbf{\bibinfo{volume}{47}},
  \bibinfo{pages}{412} (\bibinfo{year}{1982}{\natexlab{a}}).

\bibitem[{\citenamefont{Feit et~al.}(1982{\natexlab{b}})\citenamefont{Feit,
  Fleck, and Steiger}}]{bib:opsplit3}
\bibinfo{author}{\bibfnamefont{D.}~\bibnamefont{Feit}},
  \bibinfo{author}{\bibfnamefont{J.~A.} \bibnamefont{Fleck}}, \bibnamefont{and}
  \bibinfo{author}{\bibfnamefont{A.}~\bibnamefont{Steiger}},
  \bibinfo{journal}{J. Chem. Phys.} \textbf{\bibinfo{volume}{78}},
  \bibinfo{pages}{301} (\bibinfo{year}{1982}{\natexlab{b}}).

\bibitem[{\citenamefont{Takahashi and Ikeda}(1993)}]{bib:opsplit4}
\bibinfo{author}{\bibfnamefont{K.}~\bibnamefont{Takahashi}} \bibnamefont{and}
  \bibinfo{author}{\bibfnamefont{K.}~\bibnamefont{Ikeda}}, \bibinfo{journal}{J.
  Chem. Phys.} \textbf{\bibinfo{volume}{99}}, \bibinfo{pages}{8680}
  (\bibinfo{year}{1993}).

\bibitem[{\citenamefont{Gray and Manolopoulos}(1996)}]{bib:opsplit5}
\bibinfo{author}{\bibfnamefont{S.~K.} \bibnamefont{Gray}} \bibnamefont{and}
  \bibinfo{author}{\bibfnamefont{D.~E.} \bibnamefont{Manolopoulos}},
  \bibinfo{journal}{J. Chem. Phys.} \textbf{\bibinfo{volume}{104}},
  \bibinfo{pages}{7099} (\bibinfo{year}{1996}).

\bibitem[{\citenamefont{Takahashi and Ikeda}(1997)}]{bib:opsplit6}
\bibinfo{author}{\bibfnamefont{K.}~\bibnamefont{Takahashi}} \bibnamefont{and}
  \bibinfo{author}{\bibfnamefont{K.}~\bibnamefont{Ikeda}}, \bibinfo{journal}{J.
  Chem. Phys.} \textbf{\bibinfo{volume}{106}}, \bibinfo{pages}{4463}
  (\bibinfo{year}{1997}).

\bibitem[{\citenamefont{Succi and Benzi}(1993)}]{bib:qlbe1}
\bibinfo{author}{\bibfnamefont{S.}~\bibnamefont{Succi}} \bibnamefont{and}
  \bibinfo{author}{\bibfnamefont{R.}~\bibnamefont{Benzi}},
  \bibinfo{journal}{Physica D} \textbf{\bibinfo{volume}{69}},
  \bibinfo{pages}{327} (\bibinfo{year}{1993}).

\bibitem[{\citenamefont{Succi}(1996)}]{bib:qlbe2}
\bibinfo{author}{\bibfnamefont{S.}~\bibnamefont{Succi}},
  \bibinfo{journal}{Phys. Rev. E} \textbf{\bibinfo{volume}{53}},
  \bibinfo{pages}{1969} (\bibinfo{year}{1996}).

\bibitem[{\citenamefont{Succi}(2002)}]{bib:qlbe3}
\bibinfo{author}{\bibfnamefont{S.}~\bibnamefont{Succi}},
  \bibinfo{journal}{Comp. Phys. Comm.} \textbf{\bibinfo{volume}{146}},
  \bibinfo{pages}{317} (\bibinfo{year}{2002}).

\bibitem[{\citenamefont{Meyer}(1996{\natexlab{a}})}]{bib:meyer1}
\bibinfo{author}{\bibfnamefont{D.~A.} \bibnamefont{Meyer}},
  \bibinfo{journal}{Phys. Lett. A} \textbf{\bibinfo{volume}{223}},
  \bibinfo{pages}{337} (\bibinfo{year}{1996}{\natexlab{a}}).

\bibitem[{\citenamefont{Meyer}(1996{\natexlab{b}})}]{bib:meyer2}
\bibinfo{author}{\bibfnamefont{D.~A.} \bibnamefont{Meyer}},
  \bibinfo{journal}{J. Stat. Phys.} \textbf{\bibinfo{volume}{85}},
  \bibinfo{pages}{551} (\bibinfo{year}{1996}{\natexlab{b}}).

\bibitem[{\citenamefont{Meyer}(1997{\natexlab{a}})}]{bib:meyer3}
\bibinfo{author}{\bibfnamefont{D.~A.} \bibnamefont{Meyer}},
  \bibinfo{journal}{Phys. Rev. E.} \textbf{\bibinfo{volume}{55}},
  \bibinfo{pages}{5261} (\bibinfo{year}{1997}{\natexlab{a}}).

\bibitem[{\citenamefont{Meyer}(1997{\natexlab{b}})}]{bib:meyer4}
\bibinfo{author}{\bibfnamefont{D.~A.} \bibnamefont{Meyer}},
  \bibinfo{journal}{quant-ph/9712052}  (\bibinfo{year}{1997}{\natexlab{b}}).

\bibitem[{\citenamefont{Boghosian and Taylor}(1998)}]{bib:bogwash}
\bibinfo{author}{\bibfnamefont{B.~M.} \bibnamefont{Boghosian}}
  \bibnamefont{and} \bibinfo{author}{\bibfnamefont{W.}~\bibnamefont{Taylor}},
  \bibinfo{journal}{Physica D} \textbf{\bibinfo{volume}{120}},
  \bibinfo{pages}{30} (\bibinfo{year}{1998}).

\bibitem[{\citenamefont{Littlejohn}(1983)}]{bib:gyroav1}
\bibinfo{author}{\bibfnamefont{R.~J.} \bibnamefont{Littlejohn}},
  \bibinfo{journal}{J. Plasma Physics} \textbf{\bibinfo{volume}{29}},
  \bibinfo{pages}{111} (\bibinfo{year}{1983}).

\bibitem[{\citenamefont{Littlejohn}(1979)}]{bib:gyroav2}
\bibinfo{author}{\bibfnamefont{R.~J.} \bibnamefont{Littlejohn}},
  \bibinfo{journal}{J. Math. Phys.} \textbf{\bibinfo{volume}{20}},
  \bibinfo{pages}{2445} (\bibinfo{year}{1979}).

\bibitem[{\citenamefont{Boghosian}(1987)}]{bib:gyroav3}
\bibinfo{author}{\bibfnamefont{B.~M.} \bibnamefont{Boghosian}}, Ph.D. thesis,
  \bibinfo{school}{UC Davis} (\bibinfo{year}{1987}).

\bibitem[{\citenamefont{Kantor and Solodovnikov}(1973)}]{bib:quatbook}
\bibinfo{author}{\bibfnamefont{I.~L.} \bibnamefont{Kantor}} \bibnamefont{and}
  \bibinfo{author}{\bibfnamefont{A.~S.} \bibnamefont{Solodovnikov}},
  \emph{\bibinfo{title}{Hypercomplex numbers}}
  (\bibinfo{publisher}{Springer-Verlag}, \bibinfo{year}{1973}).

\end{thebibliography}
\end{document}